\documentclass{symmetry08}
\begin{document}

\FirstPageHeading{Vaneeva}
% The parameter is the label of the article. Good choice is the last name of the first author

\ShortArticleName{Reduction operators of variable coefficient semilinear
diffusion equations} % maximum 75 symbols

\ArticleName{Reduction operators of variable coefficient\\ semilinear
diffusion equations\\ with a power source}

% Names of the authors for the title of the paper
\Author{O.O. VANEEVA~$^\dag$, R.O.~POPOVYCH~$^{\dag\ddag}$ and C. SOPHOCLEOUS~$^\S$}
\AuthorNameForHeading{O.O. Vaneeva, R.O. Popovych and C. Sophocleous}
\AuthorNameForContents{VANEEVA O.O., POPOVYCH R.O. and SOPHOCLEOUS C.}
\ArticleNameForContents{Reduction\\ operators of variable coefficient semilinear
diffusion equations with\\ a power source}

\Address{$^\dag$~Institute of Mathematics of NAS of Ukraine, \\
$\phantom{{}^\dag}$\ 3 Tereshchenkivska Str., 01601 Kyiv-4, Ukraine}
\EmailD{vaneeva@imath.kiev.ua, rop@imath.kiev.ua}

\Address{$^\ddag$~Fakult\"at f\"ur Mathematik, Universit\"at Wien, \\
$\phantom{{}^\ddag}$\ Nordbergstra{\ss}e 15, A-1090 Wien, Austria}

\Address{$^\S$~Department of Mathematics and Statistics,\\
$\phantom{{}^\S}$\ University of Cyprus, Nicosia CY 1678, Cyprus}
\EmailD{christod@ucy.ac.cy}

\Abstract{Reduction operators (called often nonclassical symmetries)
 of variable-coefficient semilinear reaction--diffusion equations
with power nonlinearity $f(x)u_t=(g(x)u_x)_x+h(x)u^m$ ($m\neq0,1,2$) are investigated using the algorithm suggested
in~[O.O. Vaneeva, R.O. Popovych and C. Sophocleous, {\it Acta~Appl.~Math.}, 2009, V.106, 1--46; arXiv:0708.3457].}

\section{Introduction}
As early as in 1969 Bluman and Cole introduced a new method for finding group-invariant
(called also similarity) solutions of
partial differential equations (PDEs)~\cite{vaneeva:Bluman&Cole1969}. It was called by the authors ``non-classical''
to emphasize the difference between it and the ``classical'' Lie reduction method described, e.g., in~\cite{vaneeva:Olver1986,vaneeva:Ovsiannikov1982}.
A precise and rigorous definition of nonclassical invariance
was firstly formulated in~\cite{vaneeva:Fushchych&Tsyfra1987} as ``a generalization of the
Lie definition of invariance'' (see also~\cite{vaneeva:Zhdanov&Tsyfra&Popovych1999}).
Later operators satisfying the nonclassical invariance criterion were called, by different authors, nonclassical
symmetries, conditional symmetries and $Q$-conditional symmetries~\cite{vaneeva:FushchichSerov1988,
vaneeva:Fushchych&Shtelen&Serov&Popovych1992,vaneeva:Levi&Winternitz1989}.
Until now all names are in use. Following~\cite{vaneeva:Popovych&Vaneeva&Ivanova2007} we call nonclassical symmetries
\emph{reduction operators}. The necessary definitions, including ones of equivalence of reduction operators,
 and relevant statements on this subject are collected in~\cite{vaneeva:VPS_2009}.

The problem of finding reduction operators for a PDE is
more complicated than the similar problem on Lie symmetries because the first problem is reduced
to the integration of an overdetermined system of nonlinear PDEs, whereas in the case of Lie symmetries
one deals with a more overdetermined system of linear PDEs.
The complexity increases in times in the case of classification problem of reduction operators for
a class of PDEs having nonconstant arbitrary elements.

Often the usage of equivalence and gauging
transformations can essentially simplify the group classification problem.
Moreover, their implementation can appear to be a crucial point in solving the problem. This observation
is justified
by a number of examples~\cite{vaneeva:Ivanova&Popovych&Sophocleous2006Part1,vaneeva:VJPS2007,vaneeva:VPS_2009}.
The above transformations are of major importance for studying reduction operators
since under their classification one needs to surmount  much more essential obstacles
then those arising under the classification of Lie symmetries.

In~\cite{vaneeva:VPS_2009} simultaneous usage of equivalence transformations and mappings between classes
allowed us to carry out group classification of the class of variable coefficient
semilinear reaction--diffusion equations
with power nonlinearity
\begin{equation}\label{vaneeva:eqRDfghPower}
f(x)u_t=(g(x)u_x)_x+h(x)u^m,
\end{equation}
where $f=f(x)$, $g=g(x)$ and $h=h(x)$  are arbitrary smooth functions of the variable~$x$,
$f(x)g(x)h(x)\neq0$, $m$ is an arbitrary constant ($m\neq0,1$).

In the same paper an algorithm for finding reduction operators of class~\eqref{vaneeva:eqRDfghPower}
involving mapping between classes was proposed.
Here using this algorithm we investigate reduction operators of the equations from class~\eqref{vaneeva:eqRDfghPower}
with $m\neq2$.
The case $m=2$ will not be systematically considered since it is singular from the Lie symmetry point of view and needs
an additional mapping between classes (see~\cite{vaneeva:VPS_2009} for more details).
Nevertheless, all the reduction operators constructed for the general case $m\ne0,1,2$
are also fit for the values $m=0,1,2$.

The structure of this paper is as follows. For convenience of readers
sections 2--4 contain a short review of results obtained in~\cite{vaneeva:VPS_2009}
and used here.
Namely, in section~\ref{vaneeva:Sect_Lie_sym} all necessary information
concerning equivalence transformations and mapping of class~\eqref{vaneeva:eqRDfghPower}
to the so-called ``imaged'' class is collected.
Results on group classification and additional equivalence transformations of the imaged class are also presented.
Section 3 describes the algorithm for finding reduction operators of class~\eqref{vaneeva:eqRDfghPower} using mapping between classes.
In section~4 known reduction operators of constant-coefficient equations from the imaged  class are considered.
Their preimages are obtained.
The results of sections 5 and 6 are completely original and concern the
investigation of reduction operators  for equations from the
imaged class which have at least one nonconstant arbitrary element.
It appears that application of the reduction  method to equations from the imaged class with $m=3$ leads in some cases
to necessity of solving first-order nonlinear ODEs of a special form related to Jacobian elliptic functions.
The table with solutions of the ordinary differential equations (ODEs) of this kind is placed in Appendix.

\section{Lie symmetries and equivalence transformations}\label{vaneeva:Sect_Lie_sym}

To produce group classification of class~\eqref{vaneeva:eqRDfghPower},
it is necessary to gauge arbitrary elements of this class with equivalence transformations and
subsequent mapping of it to a simpler class~\cite{vaneeva:VPS_2009}.

\begin{theorem}\label{vaneeva:equivfgh}
The generalized extended equivalence group~$\hat G^{\sim}$ of
class~\eqref{vaneeva:eqRDfghPower} is formed by the transformations
\[\hspace{-.5\arraycolsep}
\begin{array}{l}
\tilde t=\delta_1 t+\delta_2,\quad \tilde x=\varphi(x), \quad
\tilde u=\psi(x) u, \\[1ex]
\tilde f=\dfrac{\delta_0\delta_1}{\varphi_x\psi^{2}} f, \quad
\tilde g=\dfrac{\delta_0\varphi_x}{\psi^{2}}\, g, \quad
\tilde h=\dfrac{\delta_0}{\varphi_x\psi^{m+1}} h, \quad
\tilde m=m,
\end{array}
\]
where $\varphi$ is an arbitrary smooth function of~$x$, $\varphi_x\not=0$
and $\psi$ is determined by
the formula $ \psi(x)= \bigl(\delta_3\int
\frac{dx}{g(x)}+\delta_4\bigr)^{-1}$. $\delta_j$ $(j=0,1,2,3,4)$ are arbitrary constants,
$\delta_0\delta_1({\delta_3}^2+{\delta_4}^2)\not=0$.
\end{theorem}

The usual equivalence group~$G^{\sim}$ of class~\eqref{vaneeva:eqRDfghPower} is
the subgroup of the generalized extended equivalence group~$\hat G^{\sim}$,
which is singled out with the condition $\delta_3=0$.

The presence of the arbitrary function $\varphi(x)$ in the equivalence
transformations allows us to simplify the group classification problem of
class~\eqref{vaneeva:eqRDfghPower} via reducing the number of arbitrary elements and making its more convenient for
mapping to another class.

Thus, the transformation from the equivalence group $G^{\sim}$
\begin{equation}\label{vaneeva:gauge_f=g}
t'=\mathop{\rm sign}(f(x)g(x))t,\quad
x'=\int\left|f(x)g(x)^{-1}\right|^{\frac12}\,dx, \quad
u'=u
\end{equation}
maps class~\eqref{vaneeva:eqRDfghPower} onto its subclass
$f'(x'){u'}_{t'}= (f'(x'){u'}_{x'})_{x'}+h'(x'){u'}^{m'}$
with the new arbitrary elements $m'=m,$
$
f'(x')=g'(x')=\mathop{\rm sign}(g(x))\left|f(x)g(x)\right|^{\frac12}$ and
$h'(x')=\left|g(x)f(x)^{-1}\right|^{\frac12}\,h(x).$
Without loss of generality, we can restrict ourselves to
study  the class
\begin{equation}\label{vaneeva:class_f=g}
f(x)u_t= (f(x)u_{x})_{x}+h(x)u^m,
\end{equation}
since
all results on symmetries and exact solutions for this class can be extended to
class~\eqref{vaneeva:eqRDfghPower} with transformation~\eqref{vaneeva:gauge_f=g}.

It is easy to deduce the generalized extended equivalence group for class~\eqref{vaneeva:class_f=g}
from theorem~\ref{vaneeva:equivfgh} by setting
$\tilde f=\tilde g$ and $f=g$. See theorem 4 in~\cite{vaneeva:VPS_2009}.

The next step is to make the change of the dependent variable
\begin{gather}\label{vaneeva:gauge}
v(t,x)=\sqrt{|f(x)|}u(t,x)
\end{gather}
 in class~\eqref{vaneeva:class_f=g}.
As a result, we obtain the class of related equations of the form
\begin{equation}\label{vaneeva:class_vFH}
v_t=v_{xx}+H(x)v^m+F(x)v,
\end{equation}
where the new arbitrary elements $F$ and $H$ are connected with the old ones via the formulas
\begin{gather}\label{vaneeva:FH_formulas}
F(x)=-\dfrac {(\sqrt{|f(x)|})_{xx}}{\sqrt{|f(x)|}},\quad H(x)=\dfrac {h(x)\mathop{\rm sign}f(x)}{(\sqrt{|f(x)|})^{m+1}}.
\end{gather}
Since class~\eqref{vaneeva:class_vFH} is an image of class~\eqref{vaneeva:class_f=g} with respect to
the family of transformations~\eqref{vaneeva:gauge} parameterized by the arbitrary element~$f$,
we  call them the {\it imaged class} and the {\it initial class}, respectively.

\begin{theorem}\label{vaneeva:TheoremOnGsimFH}
The generalized extended equivalence group~$\hat G^{\sim}_{FH}$ of class~\eqref{vaneeva:class_vFH}
coincides with the usual equivalence group~$G^{\sim}_{FH}$ of the same class
and is formed by the transformations
\begin{gather*}
\tilde t={\delta_1}^2 t+\delta_2,\quad \tilde x=\delta_1 x+\delta_3, \quad \tilde v=\delta_4 v, \\
\tilde F=\dfrac F{{\delta_1}^2},\quad \tilde H=\dfrac H{{\delta_1}^2{\delta_4}^{m-1}}, \quad \tilde m=m,
\end{gather*}
where $\delta_j$, $j=1,\dots,4$, are arbitrary constants, $\delta_1\delta_4\not=0$.
\end{theorem}

The following important proposition is proved in~\cite{vaneeva:VPS_2009}.
\begin{proposition}\label{vaneeva:PropositionOnEquivClassificationsUnderGaugingForClassEqRDfghPower}
The group classification in class~\eqref{vaneeva:eqRDfghPower} with respect to its
generalized extended equivalence group~$\hat G^{\sim}$ is equivalent to
the group classification in class~\eqref{vaneeva:class_vFH} with respect to
the usual equivalence group~$G^{\sim}_{FH}$ of this class.
A classification list for class~\eqref{vaneeva:eqRDfghPower} can be obtained from
a classification list for class~\eqref{vaneeva:class_vFH} by means of taking a single preimage for each
element of the latter list with respect to the resulting mapping
from class~\eqref{vaneeva:eqRDfghPower} onto class~\eqref{vaneeva:class_vFH}.
\end{proposition}

All possible $G^{\sim}_{FH}$-inequivalent values of the parameter-functions~$F$ and $H$ for which equations~\eqref{vaneeva:class_vFH}
admit  extension of Lie symmetry are listed in table~\ref{vaneeva:TableLieSymHF} together with
bases of the corresponding maximal Lie invariance algebras.
\begin{center}
\renewcommand{\arraystretch}{1.6}
\refstepcounter{table}\label{vaneeva:TableLieSymHF}
\textbf{Table~\thetable.}
The group classification of the class $v_t=v_{xx}+H(x)v^m+F(x)v$. $m\neq0,1,2$; $H(x)\neq0$.
\\[2ex]
\begin{tabular}{|c|c|c|l|}
\hline
N&$H(x)$&$F(x)$&\hfil Basis of $A^{\max}$ \\
\hline
0&$\forall$&$\forall$&$\partial_t$\\
\hline
1&
$\delta e^{qx}$&$a_1$&$\partial_t,\,\partial_x+\alpha v\partial_v$\\
\hline
2&$\delta e^{qx}$&
$-\alpha^2$&$\partial_t,\,\partial_x+\alpha v\partial_v,\,2t\partial_t+(x-2\alpha t)\partial_x+$\\ &&&
$\left(\alpha(x-2\alpha t)+\frac2{1-m}\right)v\partial_v$\\
\hline
3&$\delta x^k$&$a_2 x^{-2}$&
$\partial_t,\,2t\partial_t+x\partial_x+\frac{k+2}{1-m}\,v\partial_v$\\
\hline
4&$\delta x^ke^{px^2}$&
$-{\beta}^2x^2+a_2 x^{-2}+\gamma$&$\partial_t,\,e^{4\beta t}\partial_t+2\beta x e^{4\beta t}\partial_x -$\\&&&
$2\beta\, e^{4\beta t}\!\left(\beta x^2-\frac{k+2}{1-m}\right)v \partial_v$\\
\hline
5&$\delta e^{px^2}$&$-{\beta}^2x^2+\beta a_3$&
$\partial_t,\,e^{2\beta t}\left[\partial_x-\beta xv\partial_v\right]$\\
\hline
6&$\delta e^{px^2}$&
$-{\beta}^2x^2+\beta\frac {5-m}{1-m}$&
$\partial_t,\,e^{2\beta t}\left[\partial_x-\beta xv\partial_v\right],$\\&&&
$e^{4\beta t}\left[\partial_t+2\beta x \partial_x -
2\beta\left(\beta x^2-\frac{2}{1-m}\right)v \partial_v\right]$\\
\hline
\end{tabular}
\\[2ex]
\parbox{135mm}{Here $\alpha, \beta, \gamma, \delta, k, p, q, a_1, a_2, a_3$ are constants satisfying the conditions:
$\alpha=\tfrac q{1-m}$, $\beta=\tfrac {2p}{m-1},$ $\gamma=\beta\frac {2k+5-m}{1-m}$,
$\delta=\pm1\,{\rm mod}\,G^{\sim}_{FH}$;
$p\neq0,$ $a_1\neq-\alpha^2$, $k^2+a_2^2\neq0$, $q^2+a_1^2\neq0$,
$a_3\neq\frac{5-m}{1-m}$.}
\end{center}

The results on group classification of class~\eqref{vaneeva:class_f=g} can be found in table 3 of~\cite{vaneeva:VPS_2009}.

Additional equivalence transformations between $G^{\sim}_{FH}$-inequivalent cases of Lie symmetry extension are also constructed.
The independent pairs of point-equivalent cases from table~\ref{vaneeva:TableLieSymHF} and the corresponding transformations
are exhausted by the following:
\begin{gather}\nonumber
1\mapsto{\tilde1}|_{\smash{\tilde q=0,\,\tilde a_1=a_1+\alpha^2}},\quad 2\mapsto\tilde2|_{\tilde q=0}\colon\,\,
\tilde t=t,\quad\tilde x=x+2\alpha t,\quad \tilde v=e^{-\alpha x}v;\\[1ex]\label{vaneeva:Eq_AddEqTR4to3}
4\mapsto\tilde3\colon\,\,
\tilde t=-\frac 1{4\beta}e^{-4\beta t},\quad
\tilde x=e^{-2\beta t}x ,\quad
\tilde v=\exp{\left(\frac{\beta}{2}\,x^2+2\beta\frac{k+2}{m-1}\,t\right)}v;\\\nonumber
6\mapsto\tilde2|_{\tilde q=0}\colon\,\, \mbox{the previous transformation with}\,\, k=0.
\end{gather}
The whole set of form-preserving~\cite{vaneeva:Kingston&Sophocleous1998}
(also called admissible~\cite{vaneeva:Popovych2006c}) transformations of the imaged class
for the case $m\ne0,1,2$ is described in~\cite{vaneeva:VPS_2009}.

\section{Construction of reduction operators using\\ mappings between classes}\label{vaneeva:Sect_Algorithm}
Here we adduce the algorithm of application of equivalence transformations, gauging of arbitrary elements
and mappings between classes of equations to classification of reduction operators.

1.\ Similarly to the group classification, at first we gauge class~\eqref{vaneeva:eqRDfghPower}
to subclass~\eqref{vaneeva:class_f=g} constrained by the condition~$f=g$.
Then class~\eqref{vaneeva:class_f=g} is mapped to the imaged class~\eqref{vaneeva:class_vFH}
by transformation~\eqref{vaneeva:gauge}.

2.\ Since nonclassical symmetries of constant coefficient equations from
the imaged classes are well investigated (see
below for more details), they should be excluded from the consideration.
It also concerns variable coefficient equations from class~\eqref{vaneeva:class_vFH}
 which are point-equivalent  to
constant coefficient ones, namely equations associated with cases $1|_{q\neq0}$, $2|_{q\neq0}$ and $6$ of
table~\ref{vaneeva:TableLieSymHF} and equations reduced to them
by  transformations from the corresponding equivalence groups. As a result, only equations
from class~\eqref{vaneeva:class_vFH}  which are inequivalent with
respect to all point transformations to constant coefficient ones should be studied.

3.\ Reduction operators should be classified up to the equivalence  relations
generated by the equivalence group or even by the whole set of admissible transformations.
Only the nonsingular case $\tau\ne0$ (reduced to the case $\tau=1$) should be considered.
Operators equivalent to Lie symmetry ones should be neglected.

4.\ Preimages of the obtained nonclassical symmetries and of equations
admitting them should be found using backward gauging transformations
and mappings induced by these transformations on the sets of operators.

Reduction operators of equations from class~\eqref{vaneeva:class_f=g}
are easily found from reduction operators of corresponding equations from~\eqref{vaneeva:class_vFH}
using the formula
\begin{gather}\label{vaneeva:tr_op}
\tilde Q=\tau\partial_t+\xi\partial_x+\biggl(\frac{\eta}{\sqrt{|f|}}-\frac{\xi f_x}{2f}u\biggr)\partial_u.
\end{gather}
Here $\tau$, $\xi$ and $\eta$ are coefficients of $\partial_t$, $\partial_x$ and $\partial_v$, respectively,
in the reduction operators of equations from class~\eqref{vaneeva:class_vFH}.
The substitution $v=\sqrt{|f|}\,u$ is assumed.

There exist two ways to use mappings between classes of equations in the investigation of nonclassical symmetries.
Suppose that nonclassical symmetries of equations from the imaged class are known.
The first way is to take the preimages of both the constructed operators and the equations possessing them.
Then we can reduce the preimaged equations with respect to the corresponding preimaged operators
to find non-Lie solutions of equations from the initial class.
The above way seems to be non-optimal since the ultimate goal of the investigation of nonclassical symmetries is
the construction of exact solutions.
This observation is confirmed by the fact that the equations from the imaged class and the
associated nonclassical symmetry operators
have, as a rule, a simpler form and therefore, are more suitable than their preimages.
Reduced equations associated with equations from the imaged class are also simpler to be integrated.
Moreover, it happens that preimages of uniformly parameterized similar equations do not have similar forms and belong to
different parameterized families.
As a result, making reductions in the initial class, we have to deal with a number of different ans\"atze and reduced equations
although this is equivalent to the consideration of a single ansatz and the corresponding reduced equation within the imaged classes.
This is why the second way based on the implementation of reductions in the imaged classes and preimaging of the
obtained exact solutions instead of preimaging the corresponding reduction operators is preferable.

\section{The case of constant $\boldsymbol F$ and $\boldsymbol H$}
Constant coefficient equations from the imaged  class belong to the wider class of a
quasilinear heat equations with a source of the general form
$v_t=v_{xx}+q(v)$.
Lie and nonclassical symmetries  of these equations were investigated in~\cite{vaneeva:Dorodnitsyn1979,vaneeva:Dorodnitsyn1982}
and \cite{vaneeva:ArrigoHillBroadbridge1993,vaneeva:Clarkson&Mansfield1993,vaneeva:FushchichSerov1990,vaneeva:Serov1990}, respectively.
Their non-Lie exact solutions were constructed by the reduction method
in~\cite{vaneeva:ArrigoHillBroadbridge1993,vaneeva:Clarkson&Mansfield1993}, see also their
collection in~\cite{vaneeva:VPS_2009}.
The nonlinear equation $v_t=v_{xx}+q(v)$ possesses
pure nonclassical symmetry operators with nonvanishing coefficients of $\partial_t$
if and only if $q$ is a cubic polynomial in $v$.
Thus, in the case $q=\delta v^3+\varepsilon v$, where $\delta\ne0$, such operators
are exhausted, up to the equivalence with respect to the corresponding Lie symmetry groups, by the following:
\begin{gather*}\arraycolsep=0ex
\begin{array}{l}
\delta<0\colon\quad\partial_t\pm\frac32\sqrt{-2\delta}\,v\partial_x+\frac32(\delta v^3+\varepsilon v)\partial_v,\\[1ex]
\varepsilon=0\colon\quad\partial_t-\frac3x\partial_x-\frac{3}{x^2}v\partial_v,\\[1ex]
\varepsilon<0\colon\quad\partial_t+3\mu\tan(\mu x)\partial_x-3\mu^2 \sec^2(\mu x)v\partial_v,\\[1ex]
\varepsilon>0\colon\quad\partial_t-3\mu\tanh(\mu x)\partial_x+3\mu^2 {\rm sech}^2(\mu x)v\partial_v,\\[1ex]
\phantom{\varepsilon>0\colon\quad}\partial_t-3\mu\coth(\mu x)\partial_x-3\mu^2 {\rm cosech}^2(\mu x)v\partial_v,
\end{array}
\end{gather*}
where $\mu=\sqrt{|\varepsilon|/2}$.
Note that the last operator was missed in~\cite{vaneeva:ArrigoHillBroadbridge1993,vaneeva:Clarkson&Mansfield1993}.

Finding the preimages of equations with such values of~$q$ with respect to transformation~\eqref{vaneeva:gauge}
and the preimages of the corresponding reduction operators according to formula~\eqref{vaneeva:tr_op}, we obtain
the cases presented in table~\ref{vaneeva:TableNonclassicalSymhf_m3}.
In this table $c_1^2+c_2^2\neq0$, $\nu>0$, $\mu=\nu/\sqrt2$;
$\varepsilon=0$, $\varepsilon=\nu^2>0$ and $\varepsilon=-\nu^2<0$ in cases 1, 2 and 3, respectively.

\begin{center}\small
\renewcommand{\arraystretch}{2.2}
\refstepcounter{table}\label{vaneeva:TableNonclassicalSymhf_m3}\textbf{Table~\thetable.} Nonclassical
symmetries of equations of the form $f(x)u_t=(f(x)u_x)_x+\delta f(x)^2u^3$, $f(x)=\zeta(x)^2$.
\\[2ex]
\begin{tabular}{|c|c|l|}
\hline
N&$\zeta(x)$&\hfil Reduction operators \\
\hline
1
&$c_1x+c_2$
&$\begin{array}{l}
\partial_t\pm\frac32\sqrt{-2\delta}\zeta u\partial_x+\frac32(\delta \zeta^2 u\mp c_1\sqrt{-2\delta})u^2\partial_u,\\
\partial_t-\dfrac3x\partial_x-\dfrac{3c_2}{x^2\zeta_{_{_{}}}}u\partial_u
\end{array}$\\
\hline
2
&$\begin{array}{l}c_1\sin(\nu x)+\\[-1.5ex]c_2\cos(\nu x)\end{array}$
&$\begin{array}{l}
\partial_t\pm\frac32\sqrt{-2\delta}\zeta u\partial_x+\frac32(\delta {\zeta}^2u^2\mp\sqrt{-2\delta}\zeta_xu+\nu^2)_{}u\partial_u,
\\
\partial_t-3\mu\tanh(\mu x)\partial_x+3\mu \left(\dfrac{\zeta_x}{\zeta}\tanh(\mu x)+\mu\,{\rm sech}^2(\mu x)\right)_{_{}}u\partial_u,
\\
\partial_t-3\mu\coth(\mu x)\partial_x+3\mu \left(\dfrac{\zeta_x}{\zeta}\coth(\mu x)-\mu\,{\rm cosech}^2(\mu x)\right)_{_{}}u\partial_u\!\!\!
\end{array}$\\
\hline
3
&$\begin{array}{l}c_1\sinh(\nu x)+\\[-1.5ex]c_2\cosh(\nu x)\end{array}$
&$\begin{array}{l}
\partial_t\pm\frac32\sqrt{-2\delta}\zeta u\partial_x+\frac32(\delta {\zeta}^2u^2\mp\sqrt{-2\delta}\zeta_xu-\nu^2)u\partial_u,\\
\partial_t+3\mu\tan(\mu x)\partial_x-3\mu \left(\dfrac{\zeta_x}{\zeta}\tan(\mu x)+\mu\sec^2(\mu x)\right)_{_{}}u\partial_u
\end{array}$\\
\hline
\end{tabular}
\end{center}

\section{Reduction operators for  general values of $\boldsymbol m$}

In this section we look for $G^{\sim}_{FH}$-inequivalent reduction operators of the imaged
class~\eqref{vaneeva:class_vFH}.
Here reduction operators have the general form
$Q=\tau\partial_t+\xi\partial_x+\eta\partial_v$, where $\tau$, $\xi$ and~$\eta$ are functions of $t$,
$x$ and $v$, and $(\tau,\xi)\not =(0,0)$.
Since \eqref{vaneeva:class_vFH} is an evolution equation, there are two principally different
cases of finding $Q$: $\tau\ne0$ and $\tau=0$~\cite{vaneeva:Fushchych&Shtelen&Serov&Popovych1992,vaneeva:KunzingerPopovych,vaneeva:Zhdanov&Lahno1998}.
The singular case $\tau=0$ was exhaustively investigated for general evolution equation in~\cite{vaneeva:KunzingerPopovych,vaneeva:Zhdanov&Lahno1998}.

Consider the case $\tau\ne0$.
We can assume $\tau=1$ up to the usual equivalence of reduction operators.
Then the determining equations for the coefficients $\xi$ and $\eta$ have the form
\begin{gather}\label{vaneeva:EqDetForRedOpsOfImagedEq}\arraycolsep=0ex
\begin{array}{l}
\xi_{vv}=0,\qquad
\eta_{vv}=2(\xi_{xv}-\xi\xi_v),\\[1.5ex]
\eta_{t}-\eta_{xx}+2\xi_{x}\eta=\\[1ex]
\qquad\xi\left(H_{x}v^m+F_{x}v\right)+\left(2\xi_{x}-\eta_{v}\right)\left(H  v^m+F v\right)+\eta\left(F+
H  v^{m-1}m\right),\\[1.5ex]
3\xi_{v}\left(H v^m+F  v\right)+2\xi_{x}\xi+\xi_{t}+2\eta_{vx}-\xi
_{xx}-2\xi_{v}\eta=0.
\end{array}\end{gather}
Integration of first two equations of system~\eqref{vaneeva:EqDetForRedOpsOfImagedEq} gives us the following
expressions for $\xi$ and $\eta$
\begin{gather}\label{vaneeva:Forms_of_xi_and_eta}\arraycolsep=0ex
\begin{array}{l}
\xi=av+b,\\[1ex]
\eta=-\dfrac13a^2v^3+(a_x-ab)v^2+cv+d,
\end{array}\end{gather}
where $a=a(t,x),$ $b=b(t,x),$ $c=c(t,x)$ and $d=d(t,x).$

Substituting $\xi$ and $\eta$ from~\eqref{vaneeva:Forms_of_xi_and_eta} into the third and forth equations
of~\eqref{vaneeva:EqDetForRedOpsOfImagedEq},
 we obtain the classifying equations which include
both the residuary uncertainties in coefficients of the operator
and the arbitrary elements of the class under consideration.

Since the functions $a$, $b$, $c$, $d$, $F$ and $H$ do not depend on the variable $v$, the classifying equations
should be split with respect to different powers of $v$.

Two principally different cases $a=0$ and $a\neq0$ should be considered separately.

If $a=0$ then for any $m\neq0,1,2$ the splitting results in the system of five equations
\begin{gather}\label{vaneeva:EqDet_arbitrary_m}\arraycolsep=0ex
\begin{array}{l}
mHd=0,\quad d_t-d_{xx}+2b_xd-Fd=0,\\[1ex]
b_t-b_{xx}+2bb_x+2c_x=0,\\[1ex]
bH_x+\left(c(m-1)+2b_x\right)H=0,\\[1ex]
bF_x+2b_xF+c_{xx}-c_t-2b_xc=0.
\end{array}\end{gather}
Since $mH\neq0$ then $d=0$ and the second equation of~\eqref{vaneeva:EqDet_arbitrary_m} becomes identity.

Finding the general solution of the other three equations from~\eqref{vaneeva:EqDet_arbitrary_m} appears to be a very difficult problem.
But it is easy to construct certain particular solutions setting, e.g.,
 $b_t=0$.
This supposition implies that $c_t=0$.
Then the integration of~\eqref{vaneeva:EqDet_arbitrary_m} gives the expressions of $c$, $F$ and $H$ via the function $b(x)\neq0$
\begin{gather}\nonumber c=-\frac12b^2+\frac12b_x+k_1,\\\label{vaneeva:eq_F}
F=-\dfrac14b^2+k_1+k_2b^{-2}+b_x+\dfrac14\left(\dfrac{b_x}b\right)^2-\dfrac12\dfrac{b_{xx}}b,\\\label{vaneeva:eq_H}
H=k_3b^{-\frac{m+3}2}\exp\left[(m-1)\int\left(\dfrac b2-\dfrac{k_1}b\right)dx\right],
\end{gather}
where $k_1,$ $k_2$ and $k_3$ are arbitrary constants, $k_3\neq0$.

\begin{theorem}\label{vaneeva:theorem_general_case_of_m}
The equations from class~\eqref{vaneeva:class_vFH} with the arbitrary elements given by formulas~\eqref{vaneeva:eq_F} and~\eqref{vaneeva:eq_H}
admit reduction operators of the form
\begin{equation}\label{vaneeva:EqRedOp_general_m}
Q=\partial_t+b\partial_x+\left(-\frac12b^2+\frac12b_x+k_1\right)\!v\partial_v,
\end{equation}
where $b=b(x)$ is an arbitrary smooth function and $k_1$ is an arbitrary constant.
\end{theorem}
\begin{note}Theorem~3 is true for any $m\in\mathbb{R}$, including $m\in\{0,1,2\}$.
\end{note}
We present illustrative examples, by considering various forms of the function~$b$.
\begin{example} We take
$b=x^2$ and  substitute it in formulas~\eqref{vaneeva:eq_F}--\,\eqref{vaneeva:EqRedOp_general_m} to find that the equations
\begin{equation}\label{vaneeva:Eq_vFH_example1}
v_t=v_{xx}+\frac{k_3}{x^{m+3}}e^{\,\frac16(m-1)(x^3+6k_1x^{-1})}v^m
+\left(-\frac14x^4+\frac{k_2}{x^4}+2x+k_1\right)\!v,
\end{equation}
admit the reduction operator
\[
Q=\partial_t+x^2\partial_x+\left(-\dfrac12x^4+x+k_1\right)\!v\partial_v.
\]

The corresponding ansatz $v=xe^{k_1t-\frac 16x^3}z(\omega)$, where $\omega=t+x^{-1}$, gives the reduced ODE
\begin{equation}\label{vaneeva:Eq_Reduced_example1}
z_{\omega\omega}+k_3e^{k_1(m-1)\omega}z^m+k_2z=0.
\end{equation}

For $k_1=0$ the general solution of~\eqref{vaneeva:Eq_Reduced_example1} is written in the implicit form
\begin{gather}\label{vaneeva:eq_implicit_solution}
\int\!\left(Z-k_2z^2+C_1\right)^{-\frac12}dz=\pm\omega+C_2,\quad  Z=\begin{cases}-\frac{2k_3}{m+1}z^{m+1},&m\neq-1,\\
-2k_3\ln z,&m=-1.\end{cases}\end{gather}

If $k_2=0$ and $m\neq-1$, we are able to  integrate~\eqref{vaneeva:eq_implicit_solution}. Setting $C_1=0$,
we obtain a partial solution of the reduced equation in an explicit form:
\begin{equation}\label{vaneeva:Eq_SolOfReduced_example1}
z=\left(\pm\tfrac{m-1}2\sqrt{-\tfrac{2k_3}{m+1}}\,\omega+C\right)^\frac{2}{1-m},
\end{equation}
 where $C$ is an arbitrary constant. Note that the constant $C$ can be canceled via translations of $\omega$
induced by translations of $t$ in the initial variables.

In the case $k_1=k_2=0$ and $m\neq-1$ we construct the exact solution
 \[v= xe^{-\frac 16x^3}\!\left(\pm\tfrac{m-1}2\sqrt{-\tfrac{2k_3}{m+1}}
 \left(t+x^{-1}\right)\right)^\frac{2}{1-m}\]
of the corresponding equation~\eqref{vaneeva:Eq_vFH_example1}.
Preimages of them with respect to transformation~\eqref{vaneeva:gauge} are the equation
\begin{equation}\label{vaneeva:Eq_preimaged_example1}
e^{-\frac13x^3}x^2u_t=(e^{-\frac13x^3}x^2u_x)_x+k_3\,x^{-2}e^{-\frac13x^3}u^m
\end{equation}
and its exact solution
\[u=\left(\pm\tfrac{m-1}2\sqrt{-\tfrac{2k_3}{m+1}}\left(t+x^{-1}\right)\right)^\frac{2}{1-m}.\]

Analogously, if  $k_2=C_1=0$ and $m=-1$ then integration of~\eqref{vaneeva:eq_implicit_solution} gives
${\rm erf}\left(\sqrt{-\ln z}\right)=\pm\sqrt{\frac{2k_3}\pi}\,\omega+C,$ where ${\rm erf}(y)=\frac{2}{\sqrt{\pi}}\int_0^ye^{-t^2}dt$
is the error  function and $C$ is an arbitrary constant which can be canceled by translations of $\omega$. Therefore,
\begin{equation}\label{vaneeva:Eq_SolOfReduced_example1erf}
z=\exp\left\{{-\left[{\rm erf}^{-1}\left(\pm\sqrt{\tfrac{2k_3}\pi}\,\omega\right)\right]^2}\right\},
\end{equation}
where ${\rm erf}^{-1}$ is the inverse error function, represented by the series
\begin{gather*}{\rm erf}^{-1}y=\sum^{\infty}_{k=0}\frac{c_k}{2k+1}\left(\frac{\sqrt{\pi}}2y\right)^{2k+1},
\quad \mbox{where}\quad c_0=1,\\[1ex]
c_k=\sum^{k-1}_{m=0}\frac{c_mc_{k-1-m}}{(m+1)(2m+1)}=\left\{1,1,\frac76, \frac{127}{90},\dots\right\}.
\end{gather*}
The corresponding
exact solution of equation~\eqref{vaneeva:Eq_preimaged_example1} with $m=-1$ is
\[u=\exp\left\{-\left[{\rm erf}^{-1} \left(\pm\sqrt{\tfrac{2k_3}\pi}(t+x^{-1})\right)\right]^2\right\}.\]
\end{example}
\begin{example}
Consider $b=x^{-1}$. In view of theorem~\ref{vaneeva:theorem_general_case_of_m} the equations from class~\eqref{vaneeva:class_vFH}
with the arbitrary elements
\begin{gather}\label{vaneeva:Eq_form_FH_example2}
F=k_1+k_2x^2-2x^{-2},\quad
H=k_3x^{m+1}e^{\frac12(1-m)k_1x^2}
\end{gather}
admit the reduction operator
\[Q=\partial_t+x^{-1}\partial_x+\left(k_1-x^{-2}\right)v\partial_v.\]

The ansatz constructed with this operator is $v=x^{-1}e^{\,k_1t}z(\omega)$, where $\omega=x^2-2t$,
and the reduced equation reads
\[
4z_{\omega\omega}+k_3e^{\frac12(1-m)k_1\omega}z^m+k_2z=0.
\]

If $k_1=k_2=0$,  the reduced equation is integrated analogously to equation~\eqref{vaneeva:Eq_Reduced_example1} and
has the similar particular solution
\begin{gather*}
z=\begin{cases}\left(\pm\tfrac{m-1}2\sqrt{-\tfrac{k_3}{2(m+1)}}\,\omega\right)^\frac{2}{1-m},&m\neq-1,\\[1ex]
\exp\left\{-\left[{\rm erf}^{-1}
 \left(\pm\frac{\sqrt2}{2}\sqrt{\frac{k_3}{\pi}}\,\omega\right)\right]^2\right\},&m=-1.
 \end{cases}\end{gather*}
Substituting the obtained $z$ to the ansatz, we construct exact solutions of
equations from class~\eqref{vaneeva:class_vFH}
with the arbitrary elements~\eqref{vaneeva:Eq_form_FH_example2} for the values $k_1=k_2=0$.

The preimaged equation $x^4u_t=(x^4u_x)_x+k_3\,x^{3(m+1)}u^m$ has the exact solution
\begin{gather*}
u=\begin{cases}x^{-3}\left(\pm\tfrac{m-1}2\sqrt{-\tfrac{k_3}{2(m+1)}}\,(x^2-2t)\right)^\frac{2}{1-m},&m\neq-1,\\[1ex]
x^{-3}\exp\left\{-\left[{\rm erf}^{-1}
 \left(\pm\frac{\sqrt2}{2}\sqrt{\frac{k_3}{\pi}}\,(x^2-2t)\right)\right]^2\right\},&m=-1.
 \end{cases}\end{gather*}
\end{example}

In the following two examples we  assume that $k_1=k_2=0$ in the
formulas~\eqref{vaneeva:eq_F}--\,\eqref{vaneeva:EqRedOp_general_m} since this supposition allows us to find
preimages in class~\eqref{vaneeva:class_f=g} with arbitrary elements being elementary functions.
\begin{example}
Let $b=e^{-x}$ and $k_1=k_2=0$. The equations of the form
\begin{equation}\label{vaneeva:Eq_vFH_example3}
v_t=v_{xx}+k_3e^{\frac12\left((1-m)e^{-x}+(m+3)x\right)}v^m
-\frac14\left(e^{-2 x}+4e^{-x}+1\right)\!v,
\end{equation}
admit the reduction operator
\[
Q=\partial_t+e^{-x}\partial_x-\frac12\left(e^{-x}+e^{-2x}\right)v\partial_v.
\]

The corresponding ansatz $v=e^{\frac12(e^{-x}-x)}z(\omega)$, where $\omega=e^x-t$, gives the reduced ODE
\[
z_{\omega\omega}+k_3z^m=0.
\]
It coincides with the equation~\eqref{vaneeva:Eq_Reduced_example1} with $k_1=k_2=0,$
which has the particular solution~\eqref{vaneeva:Eq_SolOfReduced_example1}
(resp.~\eqref{vaneeva:Eq_SolOfReduced_example1erf}) for  $m\neq-1$ (resp.~$m=-1$).
Substituting these solutions to the ansatz, we obtain exact solutions of equation~\eqref{vaneeva:Eq_vFH_example3}.

A preimage of~\eqref{vaneeva:Eq_vFH_example3} with respect to transformation~\eqref{vaneeva:gauge} is the equation
\[e^{e^{-x}-x}u_t=(e^{e^{-x}-x}u_x)_x+k_3\,e^{e^{-x}+x}u^m\]
having the solution
\begin{gather*}
u=\begin{cases}\left(\pm\frac{m-1}2\sqrt{-\frac{2k_3}{m+1}}(e^x-t)\right)^\frac{2}{1-m},&m\neq-1,\\[1ex]
\exp\left\{-\left[{\rm erf}^{-1} \left(\pm\sqrt{\tfrac{2k_3}\pi}(e^x-t)\right)\right]^2\right\},&m=-1.
 \end{cases}\end{gather*}
\end{example}
\begin{example}
Substituting $b=\sin x$ and $k_1=k_2=0$ to formulas~\eqref{vaneeva:eq_F}--\eqref{vaneeva:EqRedOp_general_m}
and making the reduction procedure, we obtain the following results:
The equation
\begin{equation*}
v_t=v_{xx}+k_3(\sin x)^{-\frac12(m+3)}e^{\frac{1-m}2\cos x}v^m
+\frac14\left(\cos^2x+4\cos x+\mathop{\rm cosec}{}^2 x\right)\!v,
\end{equation*} has the exact solution
\begin{gather*}
v=\begin{cases}e^{\frac12\cos x}\sqrt{\sin x}\left(\pm\frac{m-1}2\sqrt{-\frac{2k_3}{m+1}}\left(
 t-\ln \left|\tan\frac x2\right|\right)\right)^\frac{2}{1-m},&m\neq-1,\\[1ex]
 \sqrt{\sin x}\,\exp\!\left\{\frac{\cos x}2-\left[{\rm erf}^{-1}
\left(\pm\sqrt{\tfrac{2k_3}\pi}\left(t-\ln \left|\tan\frac x2\right|\right)\right)\right]^2\right\},&m=-1.
 \end{cases}\end{gather*}
The corresponding equation from class~\eqref{vaneeva:class_f=g} is
\begin{equation*}e^{\cos x}\sin x\, u_t=(e^{\cos x}\sin x\, u_x)_x+k_3\mathop{\rm cosec}x\,e^{\cos x}u^m
\end{equation*}
whose exact solution is easy to be constructed from the above one using formula~\eqref{vaneeva:gauge}.
\end{example}

We have shown the applicability of theorem~\ref{vaneeva:theorem_general_case_of_m} for construction of non-Lie
exact solutions of equations from classes~\eqref{vaneeva:class_vFH} and~\eqref{vaneeva:class_f=g}. Moreover,
using these solutions one can find exact solutions for other equations from~\eqref{vaneeva:class_vFH}
and~\eqref{vaneeva:class_f=g} with the help of equivalence transformations from the corresponding
equivalence groups.
\begin{note} In the case $m=3$ we are able to construct more exact solutions of equations from class~\eqref{vaneeva:class_vFH}
whose coefficients are given by~\eqref{vaneeva:eq_F}--\eqref{vaneeva:eq_H} with $k_1=0$, namely, for the equations
\begin{gather}\label{vaneeva:eq_FHm3}v_t=v_{xx}+k_3b^{-3}e^{\int b\,dx}v^3+
\left(\frac{k_2}{b^2}-\dfrac14b^2+b_x+\dfrac14\left(\dfrac{b_x}b\right)^2-\dfrac12\dfrac{b_{xx}}b\right)v,
\end{gather}
where  $b=b(x)$, $k_3\neq0$.

According to theorem~\ref{vaneeva:theorem_general_case_of_m}, equation~\eqref{vaneeva:eq_FHm3} admits the
reduction operator~\eqref{vaneeva:EqRedOp_general_m} (with $k_1=0$).
 An ansatz constructed with this operator has the
form
\[v=z(\omega)\sqrt{|b|}\,e^{-\frac12\!\int b\, dx},\quad \mbox{where}\quad \omega=t-\int \frac{dx}b,\]
and reduces~\eqref{vaneeva:eq_FHm3} to the second-order ODE
\[
z_{\omega\omega}=-k_3z{}^3-k_2z.
\]
It is interesting that the reduced ODE does not depend on the function $b(x)$.
Multiplying this equation by $z_\omega$ and integrating once,  we obtain the equation
\[z_{\omega}^2=-\frac{k_3}2z{}^4-k_2z{}^2+C_1.\]
Its general solution is expressed via Jacobian elliptic functions depending on values of the
constants $k_2,$ $k_3$ and $C_1$. See Appendix for more details.

For example, if $k_2=1+\mu^2$, $k_3=-2\mu^2$ and $C_1=1$  ($0<\mu<1$) we find two exact solutions of equation~\eqref{vaneeva:eq_FHm3}
\begin{gather*}
v={\rm\mathop{sn}}\left(t-\int \frac{dx}b,\mu\right)\!\sqrt{|b|}\,e^{-\frac12\!\int b\, dx},\quad
v={\rm\mathop{cd}}\left(t-\int \frac{dx}b,\mu\right)\!\sqrt{|b|}\,e^{-\frac12\!\int b\, dx},
\end{gather*}
where ${\rm\mathop{sn}}(\omega,\mu)$, ${\rm\mathop{cd}}(\omega,\mu)$ are Jacobian elliptic functions~\cite{vaneeva:WhittakerWatson}.
\end{note}

The second case to be considered is $a\neq0$.
Then after substitution of $\xi$ and $\eta$ from~\eqref{vaneeva:Forms_of_xi_and_eta}
to system~\eqref{vaneeva:EqDetForRedOpsOfImagedEq} its last equation takes the form
\begin{gather}\label{vaneeva:Eq_m3_det}\arraycolsep=0ex
\begin{array}{l}
\dfrac23a^3v^3+2a(ab-2a_x)v^2+
(a_t+3a_{xx}+3aF-2(ab)_x-2ac)v\,+\\[1ex]
\qquad b_t+2b_xb-b_{xx}-2ad+2c_x+3aHv^m=0.
\end{array}\end{gather}
It is easy to see that $a\neq0$ if and only if $m=3$.
The investigation of this case is the subject of the next section.

\section{Specific reduction operators for the cubic\\ nonlinearity}

Splitting equation~\eqref{vaneeva:Eq_m3_det} in the case $m=3$ and $a\neq0$ with respect to $u$,
 we obtain that the
functions $a$, $b$ $c$ and $d$ do not depend on the variable $t$ and are expressed via the functions $F$ and $H$
in the following way
\begin{gather}\label{vaneeva:Eq_bcd}\arraycolsep=0ex
\begin{array}{l}
a=\dfrac32\sqrt{2}\,\varepsilon\sqrt{-H},\quad b=\dfrac{H_x}{H},\quad
c=\dfrac18\left
(12F-2\left(\dfrac{H_x}H\right)_x-\left(\dfrac{H_x}H\right)^2\right),\\[2ex]
d=%\dfrac{\sqrt {2}\,\varepsilon}4\,{\dfrac { 3\,{H_{{x}}}^{3}-4\,HH_xH_{xx}+{H}^{2}H_{xxx}-2\,F_{{x}}{H}^{3}  }{ \left( -H \right) ^{7/2}}}=\\
\dfrac{\sqrt {2}\,\varepsilon}{2\sqrt{-H}}\left(F_x+\dfrac12\dfrac{H_x}H\left(\dfrac{H_x}H\right)_x-
\dfrac12\left(\dfrac{H_x}H\right)_{xx}\right),
\end{array}\end{gather}
 where $\varepsilon=\pm1$. If $H<0$ the corresponding reduction operators have real coefficients.

Then splitting of the third equation of system~\eqref{vaneeva:EqDetForRedOpsOfImagedEq}
 for $m=3$  results in the system of two ordinary differential equations
\begin{gather}\label{vaneeva:system_m=3}\arraycolsep=0ex
\begin{array}{l}
H^3H_{xxxx}-13\,H^4_x+2\,F_xH^3H_x +22\,HH^2_xH_{xx}
 -4\,FH^2H^2_x-\\[1ex]4\,H^2H^2_{xx}-6\,H^2H_xH_{xxx}+
 4\,FH^3H_{xx}-6\,F_{xx}H^4=0,\\[2ex]
16F_{xxx}H^5+16\,H^2H_x H^2_{xx}+3\,H^2H^2_x H_{xxx}-4\,F_xH^4 H_{xx}
 -\\[1ex]6\,H^3H_{xx} H_{xxx}
  -18\,HH^3_x H_{xx}-
  8\,FF_xH^5+2\,F_xH^3H^2_x -\\[1ex]20\,FH^2H^3_x
-12\,F H^4H_{xxx}+5\,H^5_x+32\,F H^3 H_x H_{xx}=0.
\end{array}\end{gather}

The following statement is true.
\begin{theorem}\label{vaneeva:theorem_RedOp_m=3}
The equations from class~\eqref{vaneeva:class_vFH} with $m=3$ and the arbitrary elements
satisfying system~\eqref{vaneeva:system_m=3}
admit reduction operators of the form
\begin{gather}\label{vaneeva:Eq_m=3_RedOp}\arraycolsep=0ex
\begin{array}{l}
Q=\partial_t+\left(\dfrac32\sqrt{2}\,\varepsilon\sqrt{-H}\,v+\dfrac{H_x}{H}\right)\partial_x\,+\\[1.5ex]
\phantom{Q=}\Biggl[\dfrac32H v^3+\dfrac34\sqrt{2}\,\varepsilon\dfrac{H_x}{\sqrt{-H}}\,v^2+\dfrac18\left
(12F-2\left(\dfrac{H_x}H\right)_x-\left(\dfrac{H_x}H\right)^2\right)v\,+\\[1.5ex]
\phantom{Q=\Biggl[}\dfrac{\sqrt {2}\,\varepsilon}{2\sqrt{-H}}\left(F_x+\dfrac12\dfrac{H_x}H\left(\dfrac{H_x}H\right)_x-
\dfrac12\left(\dfrac{H_x}H\right)_{xx}\right)\Biggr]\partial_v,
\end{array}\end{gather}
where $\varepsilon=\pm1$.
\end{theorem}

Let us note that system~\eqref{vaneeva:system_m=3} can be rewritten in the simpler form in terms of the
functions $F$ and $b$
\begin{gather}\label{vaneeva:system_m=3b}\arraycolsep=0ex
\begin{array}{l}
\phantom{16\,\,}b_{{xxx}}=6\,F_{{xx}}-2
\,b_{{x}}{b}^{2}+{b_{{x}}}^{2}+2\,bb_{{xx}}-2\,F_{{x}}b-4\,Fb_{{x}},\\[1.5ex]
16\,F_{{xxx}}=4\,b_{{x}}F_{{x}}+2\,{b_{{x}}}^{2}b+6\,b_{{x}}b_{{xx}}+2\,{b}^{2}F_{{x}}+{b}^{3}b_{{x}}+\\[1ex]
\phantom{16\,F_{{xxx}}=\,\,}3\,{b}^{2}b_{{xx}}+12\,Fb_{{xx}}+8\,FF_{{x}}+4\,Fbb_{{x}}.
\end{array}\end{gather}

System~\eqref{vaneeva:system_m=3} consists of two nonlinear fourth- and third-order ODEs.
Unfortunately we were not able to find its general solution. Nevertheless,
 we tested the six pairs of functions $F$ and $H$ appearing
in table~1 in order to check whether they satisfy system~\eqref{vaneeva:system_m=3}. In the case of positive
answer the corresponding reduction operator is easily constructed via formula~\eqref{vaneeva:Eq_m=3_RedOp}.
It appears that system~\eqref{vaneeva:system_m=3} is satisfied by $F$ and $H$ from cases 1, 2 and 6 and by those
from cases 3 and 4 for special values of the constants $k$ and $a_2$, namely, $(k, a_2)\in\left
\{\left(-3,\tfrac94\right),\,\left(-\tfrac32,\tfrac3{16}\right)\right\}$.

So, we can construct preimages of these equations using formulas~\eqref{vaneeva:FH_formulas}.
Below we list the pairs of the coefficients $f$ and $h$ for which the corresponding equations from class~\eqref{vaneeva:class_f=g}
with $m=3$ admit nontrivial reduction operators.

Hereafter $b_1^2+b_2^2\neq0$. The numbers of cases coincide with the numbers of the corresponding cases from table~1.
(Case 5 does not appear below since the functions $F$ and $H$ from this case of table 1 do not satisfy
system~\eqref{vaneeva:system_m=3}.)
\begin{gather*}\boldsymbol{1}.\,\,
a_1=0\colon\qquad f=(b_1x+b_2)^2,\quad h=\delta e^{qx}(b_1x+b_2)^4,\quad q\neq0;\\
\phantom{\boldsymbol{1}.\,\,}a_1>0\Rightarrow a_1=1 \bmod G^{\sim}_{FH}\colon\quad f=(b_1\sin x+b_2\cos x)^2,\\
\phantom{\boldsymbol{1}.\,\,a_1>0\Rightarrow a_1=1 \bmod G^{\sim}_{FH}\colon\quad }h=\delta e^{qx}(b_1\sin x+b_2\cos x)^4;\\
\phantom{\boldsymbol{1}.\,\,}a_1<0\Rightarrow a_1=-1\bmod G^{\sim}_{FH}\colon\,f=(b_1\sinh x+b_2\cosh x)^2,\\
\phantom{\boldsymbol{1}.\,\, a_1<0\Rightarrow a_1=-1\bmod G^{\sim}_{FH}\colon\,}h=\delta e^{qx}(b_1\sinh x+b_2\cosh x)^4,\,\,
 q\neq\pm2.\\[2ex]
\boldsymbol{2}.\,\,q=0\colon\qquad f=(b_1x+b_2)^2,\quad h=\delta(b_1x+b_2)^4;\\
\phantom{\boldsymbol{2}.\,\,}q\neq0\Rightarrow q=-2\bmod G^{\sim}_{FH}\colon\quad f=(b_1\sinh x+b_2\cosh x)^2,\\
\phantom{\boldsymbol{2}.\,\,q\neq0\Rightarrow q=-2\bmod G^{\sim}_{FH}\colon\quad}h=\delta e^{-2x}(b_1\sinh x
+b_2\cosh x)^4.\\[2ex]
\boldsymbol{3}.\,\,(k, a_2)=\left(-3,\tfrac94\right)\colon\, f=x(b_1\sin(\sqrt{2} \ln|x|)+b_2\cos(\sqrt{2} \ln|x|))^2,\\
\phantom{\boldsymbol{3}.\,\,(k, a_2)=\left(-3,\tfrac94\right))\colon\, }
h=\delta x^{-1}(b_1\sin(\sqrt{2} \ln|x|)+b_2\cos(\sqrt{2} \ln|x|))^4;\\
\phantom{\boldsymbol{3}.\,\,}(k, a_2)=\left(-\tfrac32,\tfrac3{16}\right)\colon\,f=x(b_1|x|^{\frac14}+b_2|x|^{-\frac14})^2,\\
\phantom{\boldsymbol{3}.\,\,(k, a_2)=\left(-\tfrac32,\tfrac3{16}\right)\colon\,}
h=\delta \sqrt{|x|}(b_1|x|^{\frac14}+b_2|x|^{-\frac14})^4.\\[2ex]
\boldsymbol{4}.\,\,f=x^{-1}\left(b_1 M_{\kappa,\mu}({p} x^2)+b_2 W_{\kappa,\mu}({p} x^2)\right)^2,\\
\phantom{\boldsymbol{4}.\,\,}h=\delta x^{k-2} e^{px^2}\left(b_1 M_{\kappa,\mu}({p} x^2)+b_2 W_{\kappa,\mu}({p} x^2)\right)^4,
\end{gather*}
where  $\kappa=-\frac{k+1}{4}$, $\mu=\frac{\sqrt{|1-4a_2|}}4$, $(k, a_2)\in\left\{\left(-3,\tfrac94\right),\,
\left(-\tfrac32,\tfrac3{16}\right)\right\} $.  $M_{\kappa,\mu}$ and $W_{\kappa,\mu}$ are the Whittaker functions~\cite{vaneeva:WhittakerWatson}.
\begin{gather*}
\boldsymbol{6}.\,\,f=x^{-1}\left(b_1 M_{-\frac14,\frac14}({p} x^2)+b_2 W_{-\frac14,\frac14}({p} x^2)\right)^2,\\
\phantom{\boldsymbol{6}.\,\,}h=\delta
x^{-2} e^{px^2}\left(b_1 M_{-\frac14,\frac14}({p} x^2)+b_2 W_{-\frac14,\frac14}({p} x^2)\right)^4.
\end{gather*}
Note that in the case $p>0$ the above Whittaker function is expressed via the error function:
 $M_{-\frac14,\frac14}(p x^2)=\frac12\,\sqrt\pi\,\sqrt[4]{px^2}\,e^{\frac p2\,x^2}{\rm erf}( \sqrt{px^2})$~\cite{vaneeva:AbramowitzStegun}.

Since the cases
$1|_{q\neq0}$, $2|_{q\neq0}$ and $6$ are reduced to constant-coefficient ones we do not consider them.
\begin{example}
Class~\eqref{vaneeva:class_vFH} contains equations with cubic nonlinearity, which are not reduced to constant-coefficient ones
by point transformations and admit reduction operators of the form~\eqref{vaneeva:Eq_m=3_RedOp}.
One of them is the equation with the coefficients $F$ and $H$ presented by case 3 of table~\ref{vaneeva:TableLieSymHF}
with $k=-3,$ $a_2=\frac94$ and $\delta=-1$, namely,
\begin{equation}\label{vaneeva:Eq_vFH_example5}
v_t=v_{xx}-x^{-3}v^3+\frac94x^{-2}v.
\end{equation}
According to theorem~\ref{vaneeva:theorem_RedOp_m=3} this equation admits two similar reduction operators ($\varepsilon=\pm1$)
\begin{gather*}
Q_{\pm}=\partial_t+\frac32\sqrt 2\left(\varepsilon x^{-\frac32}v-\sqrt 2\,x^{-1}\right)\!\partial_x\,-\\
\phantom{Q=}\quad\frac34\sqrt 2\left(\sqrt 2\,x^{-3}v^3-3
\varepsilon x^{-\frac52}v^2-\sqrt 2\,x^{-2}v+4\varepsilon x^{-\frac32}\right)\!\partial_v.
\end{gather*}
They lead to the solutions differing only in their signs.
Since equation~\eqref{vaneeva:Eq_vFH_example5} is invariant with respect to the transformation  $v\mapsto -v$,
we consider in detail only the case $\varepsilon=1$.
For all expressions to be correctly defined, we have to restrict ourself with values $x>0$.
(Another way is to replace $x$ by $|x|$.)

For convenient reduction we apply the hodograph transformation
\[\tilde t=v,\quad\tilde x=x,\quad\tilde v=t\]
which maps equation~\eqref{vaneeva:Eq_vFH_example5} and the reduction operator~$Q_{+}$ to the equation
\begin{equation}\label{vaneeva:Eq_vFH_example5_hod}
{\tilde v}_{\tilde t}{}^2\,{\tilde v}_{\tilde x\tilde x}+{\tilde v}_{\tilde x}{}^2\,{\tilde v}_{\tilde t\tilde t}-
2\,{\tilde v}_{\tilde t}\,{\tilde v}_{\tilde x}\,{\tilde v}_{\tilde t\tilde x}+{\tilde v}_{\tilde t}{}^2
+\frac{{\tilde t}^{\,3}}{{\tilde x}^3}\,{\tilde v}_{\tilde t}{}^3-
\frac94\frac{\tilde t}{{\tilde x}^2}\,{\tilde v}_{\tilde t}{}^3=0
\end{equation}
and its reduction operator
\begin{gather*}
\tilde Q_{+}=-\frac34\sqrt 2\left(\sqrt 2\,{\tilde x}^{-3}{\tilde t}^{\,3}-3
{\tilde x}^{-\frac52}{\tilde t}^{\,2}-\sqrt 2\,{\tilde x}^{-2}{\tilde t}
+4{\tilde x}^{-\frac32}\right)\!\partial_{\tilde t}\,+\\
\phantom{Q=}\quad\frac32\sqrt 2\left({\tilde x}^{-\frac32}\tilde t-
\sqrt 2\,{\tilde x}^{-1}\right)\!\partial_{\tilde x}+\partial_{\tilde v},
\end{gather*}
respectively.
An ansatz constructed with the operator~$\tilde Q_{+}$ has the form
\[
\tilde v=\frac1{24}{\tilde x}^2\frac{\tilde t+\sqrt{2\tilde x}}{\tilde t-\sqrt{2\tilde x}}-\frac1{12}{\tilde x}^2+
z(\omega),\quad\mbox{where}\quad\omega={\tilde x}^2\frac{\tilde t-\sqrt{2\tilde x}}{\tilde t+\sqrt{2\tilde x}},
\]
and reduces~\eqref{vaneeva:Eq_vFH_example5_hod} to the simple linear ODE
$\omega z_{\omega\omega}+2z_{\omega}=0$ whose general solution $z=\tilde c_1+\tilde c_2\omega^{-1}$
substituted to the ansatz gives the exact solution
\[
\tilde v=\frac{{\tilde x}^4+24\tilde c_2}{24{\tilde x}^2}\frac{\tilde t+\sqrt{2\tilde x}}{\tilde t-\sqrt{2\tilde x}}-\frac1{12}{\tilde x}^2+
\tilde c_1
\]
of equation~\eqref{vaneeva:Eq_vFH_example5_hod}.  Applying
 the inverse hodograph transformation and canceling the constant $\tilde c_1$ by translations with respect to~$t$,
we construct the non-Lie solution
\begin{equation}\label{vaneeva:Eq_vFH_example5_Solution}
v=\sqrt{2x}\,\frac{3x^4+24tx^2+c_2}{x^4+24tx^2-c_2}
\end{equation}
of equation~\eqref{vaneeva:Eq_vFH_example5}.
The solution \eqref{vaneeva:Eq_vFH_example5_Solution} with $c_2=0$ is a Lie solution invariant with respect to
the dilatation operator $D=4t\partial_t+2x\partial_x+\partial_v$ from the maximal Lie invariance algebra
of equation~\eqref{vaneeva:Eq_vFH_example5}.
However, it is much harder to find this solution by the reduction with respect to the operator $D$.
The corresponding ansatz $v=\sqrt{x}z(\omega)$, where $\omega=t^{-1}x^2$, has a simple form
but the reduced ODE $4\omega{}^2z_{\omega\omega}+\omega(\omega+4)z_{\omega}+2z-z^3=0$
is nonlinear and complicated.

This example justifies the observation made by W.~Fushchych~\cite{vaneeva:Fushchich1991} that
``ansatzes generated by conditional symmetry operators often reduce an initial nonlinear equation to a linear one.
As a rule, a Lie reduction does not change the nonlinear structure of an equation.''
We can also formulate the more general similar observation that
a complicated non-Lie ansatz may lead to a simple reduced equation while
a simple Lie ansatz may give a complicated reduced equation which is difficult to be integrated.

One of the preimages of equation~\eqref{vaneeva:Eq_vFH_example5} with respect to
transformation~\eqref{vaneeva:gauge} is the equation
 \[
x\sin^2(\sqrt{2} \ln x)u_t=
\left(x\sin^2(\sqrt{2} \ln x)u_x\right)_x-x^{-1}\sin^4(\sqrt{2} \ln x)\,u^3,
\]
having the non-Lie exact solution
\[u=\sqrt{\frac2x}\,|\mathop{\rm cosec}(\sqrt{2} \ln x)|\frac{3x^4+24tx^2+c_2}{x^4+24tx^2-c_2}.\]
\end{example}
\begin{example}
Consider the equation from the imaged class~\eqref{vaneeva:class_vFH}
\begin{equation}\label{vaneeva:Eq_vFH_example6}
v_t=v_{xx}-x^{-\frac32}v^3+\frac3{16}\frac{v}{x^2}
\end{equation}
for the values $x>0$
(case 3 of table~\ref{vaneeva:TableLieSymHF}
with $m=3$, $k=-\frac32,$ $a_2=\frac3{16}$ and $\delta=-1$).
It admits the reduction operator of form~\eqref{vaneeva:Eq_m=3_RedOp}
\begin{gather*}
Q_{+}=\partial_t+\frac32\left({\sqrt{2}}\,x^{-\frac34}v-x^{-1}\right)\partial_x-
\frac38\left(4 x^{-\frac32}v^3-3
\,{\sqrt{2}} x^{-\frac74}v^2+x^{-2}v\right)\partial_v.
\end{gather*}
Usage of the same technique as in the previous example gives the non-Lie exact solution
of~\eqref{vaneeva:Eq_vFH_example6}
\begin{equation}\label{vaneeva:Eq_vFH_example6_Solution}
v={\frac12{5\sqrt2}\, x^\frac14 \frac { 3\,t+x^2}{\sqrt{x}(15\,t
+x^2)+c_2}}.
\end{equation}
Applying the transformation~$v=\sqrt{x}(b_1x^{\frac14}+b_2x^{-\frac14})\,u$ to solution~\eqref{vaneeva:Eq_vFH_example6_Solution},
we obtain a non-Lie solution of the equation
\begin{gather*}x(b_1x^{\frac14}+b_2x^{-\frac14})^2u_t=\\\qquad\qquad
\left(x(b_1x^{\frac14}+b_2x^{-\frac14})^2u_x\right)_x
-\sqrt{x}(b_1x^{\frac14}+b_2x^{-\frac14})^4u^3
\end{gather*}
from class~\eqref{vaneeva:class_f=g}, where $b_1$ and $b_2$ are arbitrary constants, $b_1^2+b_2^2\neq0$.
\end{example}

The equivalence of cases 3 and 4 from table 1 with respect to the point
transformation~\eqref{vaneeva:Eq_AddEqTR4to3}
allows us to use
solutions~\eqref{vaneeva:Eq_vFH_example5_Solution} and~\eqref{vaneeva:Eq_vFH_example6_Solution}
for finding non-Lie exact solutions of the equations
\begin{gather}\label{vaneeva:eq_case4_1}v_t=v_{xx}-x^{-3}e^{px^2}v^3+\left(-p^2x^2+\tfrac94x^{-2}+2p\right)v,\\\label{vaneeva:eq_case4_2}
v_t=v_{xx}-x^{-\frac32}e^{px^2}v^3+\left(-p^2x^2+\tfrac3{16}x^{-2}+\tfrac p2\right)v\end{gather}
(case 4 of table~1, where $m=3$, $(k, a_2)\in\bigl\{\left(-3,\tfrac94\right),\,\left(-\tfrac32,\tfrac3{16}\right)\bigr\}$ and $\delta=-1$).
The obtained solutions of~\eqref{vaneeva:eq_case4_1} and~\eqref{vaneeva:eq_case4_2} are respectively
\begin{gather*}
v=e^{-\frac p2 x^2}\sqrt{2x}\,\dfrac{3px^4-6x^2+c_2e^{8pt}}{px^4-6x^2-c_2e^{8pt}}\\[-.5ex]\hspace*{-9mm}\mbox{and}\\[-1ex]
v={{5\sqrt2}\,e^{-\frac p2 x^2} x^\frac14 \dfrac { 4px^2-3}{2\sqrt{x}(4px^2-15)+c_2e^{5pt}}}.
\end{gather*}

\section{Appendix}
After reducing an equation from class~\eqref{vaneeva:class_vFH} with the coefficients given
by~\eqref{vaneeva:eq_F},~\eqref{vaneeva:eq_H} ($m=3,$ $k_1=0$) by means of operator~\eqref{vaneeva:EqRedOp_general_m},
we need to integrate an ODE of the form  $z_{\omega}^2=Pz^4+Qz^2+R$, where $P$, $Q$ and $R$ are real constants
(see note~1). By scale transformations, this equation can be transformed to one from those with righthand sides
adduced in the fourth column of table~\ref{vaneeva:TableEllipticFunctions}.
The corresponding solutions are Jacobian  elliptic functions~\cite{vaneeva:AbramowitzStegun,vaneeva:WhittakerWatson}.
Below
\begin{alignat*}{3}
&{\rm\mathop{cd}}(\omega,\mu)=\dfrac{{\rm\mathop{cn}}(\omega,\mu)}{{\rm\mathop{dn}}(\omega,\mu)},\quad
&&{\rm\mathop{ns}}(\omega,\mu)=\dfrac1{{\rm\mathop{sn}}(\omega,\mu)},\quad
&&{\rm\mathop{dc}}(\omega,\mu)=\dfrac1{{\rm\mathop{cd}}(\omega,\mu)},\\
&{\rm\mathop{nc}}(\omega,\mu)=\dfrac1{{\rm\mathop{cn}}(\omega,\mu)},\quad
&&{\rm\mathop{nd}}(\omega,\mu)=\dfrac1{{\rm\mathop{dn}}(\omega,\mu)},\quad
&&{\rm\mathop{sc}}(\omega,\mu)=\dfrac{{\rm\mathop{sn}}(\omega,\mu)}{{\rm\mathop{cn}}(\omega,\mu)},\\
&{\rm\mathop{sd}}(\omega,\mu)=\dfrac{{\rm\mathop{sn}}(\omega,\mu)}{{\rm\mathop{dn}}(\omega,\mu)},\quad
&&{\rm\mathop{cs}}(\omega,\mu)=\dfrac1{{\rm\mathop{sc}}(\omega,\mu)},\quad
&&{\rm\mathop{ds}}(\omega,\mu)=\dfrac1{{\rm\mathop{sd}}(\omega,\mu)}.
\end{alignat*}
The parameter $\mu$ is a real number. Without loss of generality $\mu$ is supposed to be in the closed interval $[0,1]$
since elliptic functions whose parameter is real can be made to depend on elliptic functions whose parameter lies
between 0 and~1 \cite[\S 16]{vaneeva:AbramowitzStegun}. If $\mu$ is equal to 0 or 1, the
Jacobian elliptic functions degenerate to elementary ones.

\begin{center}%\small
\renewcommand{\arraystretch}{1.5}
\refstepcounter{table}\label{vaneeva:TableEllipticFunctions}
\textbf{Table~\thetable~(\!\cite{vaneeva:WangZhou}).}
Relations between values of $(P,Q,R)$ and corresponding solutions $z(\omega)$ of the ODE
$z_{\omega}^2=Pz^4+Qz^2+R$.
\\[2ex]
\begin{tabular}{|c|c|c|c|c|}
\hline
$P$&$Q$&$R$&$P{z}^4+Q{z}^2+R$&${z}(\omega)$ \\
\hline
$\mu^2$&$-(1+\mu^2)$&$1$&$\left(1-{z}^2\right)\left(1-\mu^2{z}^2\right)$&${\rm\mathop{sn}}(\omega,\mu),$\\
&&&&${\rm\mathop{cd}}(\omega,\mu)$\\
\hline
$-\mu^2$&$2\mu^2-1$&$1-\mu^2$&$\left(1-{z}^2\right)\left(\mu^2{z}^2+1-\mu^2\right)$&${\rm\mathop{cn}}(\omega,\mu)$\\
\hline
$-1$&$2-\mu^2$&$\mu^2-1$&$\left(1-{z}^2\right)\left({z}^2+\mu^2-1\right)$&${\rm\mathop{dn}}(\omega,\mu)$\\
\hline
$1$&$-(1+\mu^2)$&$\mu^2$&$\left(1-{z}^2\right)\left(\mu^2-{z}^2\right)$&${\rm\mathop{ns}}(\omega,\mu),$\\
&&&&${\rm\mathop{dc}}(\omega,\mu)$\\
\hline
$1-\mu^2$&$2\mu^2-1$&$-\mu^2$&$\left(1-{z}^2\right)\left((\mu^2-1){z}^2-\mu^2\right)$&${\rm\mathop{nc}}(\omega,\mu)$\\
\hline
$\mu^2-1$&$2-\mu^2$&$-1$&$\left(1-{z}^2\right)\left((1-\mu^2){z}^2-1\right)$&${\rm\mathop{nd}}(\omega,\mu)$\\
\hline
$1-\mu^2$&$2-\mu^2$&$1$&$\left(1+{z}^2\right)\left((1-\mu^2){z}^2+1\right)$&${\rm\mathop{sc}}(\omega,\mu)$\\
\hline
$\mu^2(\mu^2-1)$&$2\mu^2-1$&$1$&$\left(1+\mu^2{z}^2\right)\left((\mu^2-1){z}^2+1\right)$&${\rm\mathop{sd}}(\omega,\mu)$\\
\hline
$1$&$2-\mu^2$&$1-\mu^2$&$\left(1+{z}^2\right)\left({z}^2+1-\mu^2\right)$&${\rm\mathop{cs}}(\omega,\mu)$\\
\hline
$1$&$2\mu^2-1$&$\mu^2(\mu^2-1)$&$\left({z}^2+\mu^2\right)\left({z}^2+\mu^2-1\right)$&${\rm\mathop{ds}}(\omega,\mu)$\\
\hline
\end{tabular}
\end{center}

\subsection*{Acknowledgements}
The research of ROP was supported  by the Austrian Science Fund (FWF),  project P20632. OOV and ROP
are grateful for the hospitality provided by the University of Cyprus.

{\footnotesize

$^*$\phantom{$^*$}Available at http://www.imath.kiev.ua/$\sim$\,fushchych

\noindent
$^{**}$Available at http://www.imath.kiev.ua/$\sim$\,appmath/Collections/collection2006.pdf
}\LastPageEnding
\end{document}